# Resonance Microwave Absorption in He II.

A.S.Rybalko, S.P.Rubets, E.Ya.Rudavskii, V.A.Tikhiy, S.I.Tarapov*, R.V.Golovashchenko*, and V.N.Derkach*

B. Verkin Institute for Low Temperature Physics and Engineering National Academy of Science of Ukraine, 47 Lenin Ave., 61103 Kharkov, Ukraine

*Institute of Radiophysics and Electronics National Academy of Science of Ukraine, 12 Proskury Str., 61085 Kharkov, Ukraine

rybalko@ilt.kharkov.ua

**Abstract**

Microwave (MW) absorption in liquid $^4$He is investigated in the frequency range of 40-200 GHz at T = 1.4 - 2.5 K. "Whispering gallery" of waves was generated by a dielectric disc resonator immersed into the liquid. Resonance absorption of MWs was detected at 180.3 GHz, which corresponds to the roton minimum of the liquid helium excitation spectrum. The creation of a single roton is possible because of the presence of the resonator wall which absorbs an extra momentum. The resonance frequency is shown to decreases with temperature in an excellent agreement with the temperature dependence of the roton gap obtained previously in the neutron scattering experiment. The temperature dependence of the MW absorption data indicates the anomalous behavior near the λ-point and displays the hysteretic character.

1. **Introduction**

An observation of electric induction in superfluid $^4$He induced by relative motion of its normal and superfluid components has been recently reported [1, 2]. This new extraordinary effect was revealed in two different types of experiments: electric potential oscillations were generated by a second-sound wave of which the velocities of normal $\upsilon_n$ and superfluid $\upsilon_s$ components relate as $\upsilon_n = \upsilon_s \frac{\rho_s}{\rho_n}$ (here $\rho_s$ and $\rho_n$ are the densities of the superfluid and normal components, respectively), and by liquid helium motion in a torsion oscillator whose walls affect the normal component only. In both experiments, the amplitude of the electric potential induced was about $10^{-8} - 10^{-7}$ V. For liquid helium consisting of electrically neutral and spherically symmetrical atoms, this observation was not expected. A reverse effect (second-sound wave generation by an ac electric field) was also observed.

The experimental finding of mechanoelectric effect in He II has stimulated a number of theoretical studies aiming to explain its mechanism [3-8]. On the basis of experiments [1, 2], it could be assumed that the relative motion of the normal and superfluid components is a result of internal electromagnetic forces related to the macroscopic quantum ordering. Anyway, these experiments indicate that there is a certain relationship between mechanical and electric phenomena in superfluid helium whose nature is not clearly understood.

In this work, we continue our study the relationship between mechanical and electric phenomena in superfluid helium using microwaves. It is expected that the electric component of a MW will generate superfluid and normal flows in He II which, in turn, will interact with the MW. In addition to previous studies of MW absorption in He II [9-11], we use microwaves of substantially higher frequencies comparable with typical frequencies of rotons.

## 2. Experiment.

The setup used to measure the absorption of an electromagnetic wave by liquid helium in the frequency range of 40 – 198 GHz is shown schematically in Fig. 1.

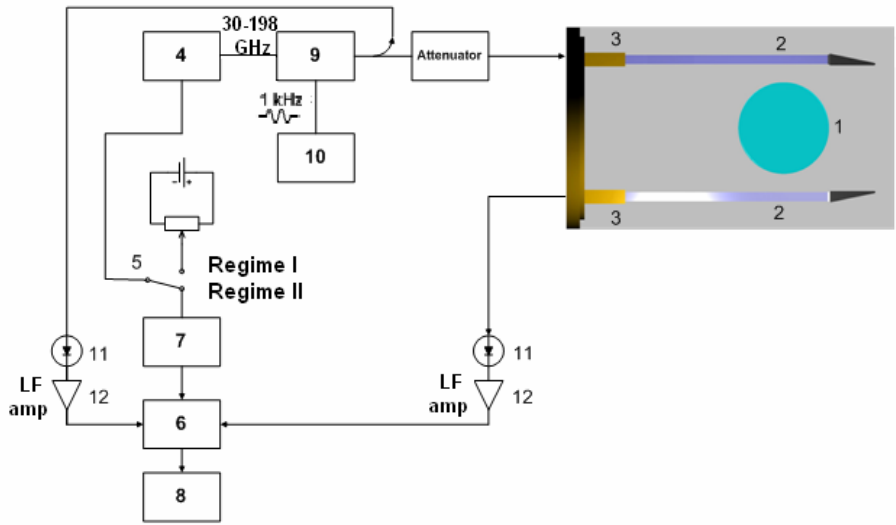

Fig. 1. Diagram of experiment:

1 – dielectric disc resonator,
2 – dielectric waveguides,
3 – hollow metallic waveguide,
4 – electromagnetic source,
5 – operation regime switch,
6 – detecting module,
7 – matching block,
8 – personal computer,
9 - modulator,
10 – low-frequency oscillator,
11 - detector,
12 - amplifier.

The measurement was made with a dielectric disc resonator operating in the "whispering gallery" mode [12-14]. Resonator 1 with $d$ = 19.00 mm in diameter and $h$ = 1.0 mm in thickness was perfectly suitable for generating high-Q whispering gallery modes [13]:

$$d \geq \frac{m\lambda_0 \sqrt{3}}{\pi\sqrt{\varepsilon}}, \quad h \leq \frac{\lambda_0}{\sqrt{\varepsilon}}, \tag{1}$$

where $\lambda$ is the vacuum space wavelength, $m$ is the azimuthal mode index, $\varepsilon$ is the material dielectric permittivity. Resonator 1 was manufactured with the optical precision high-quality crystalline quartz. Its principal axis was along that of the cylinder. Prior to the low temperature experiment, the resonator surface was carefully cleaned in an ultrasonic bath.

It is known that the device measuring the absorption of electromagnetic waves must be perfectly matched with the system under investigation to have comparable wave resistance. The Q-factor of the resonator and hence its wave resistance $R_{oe}$ increases considerably at low temperatures. It was therefore important to restrict the growth of $R_{oe}$. For this purpose the distance between antennas 2 and resonator 1 was selected so that the waveguides could induce additional losses thereby limiting the resonator Q-factor in vacuum to $Q \sim 10^5$. In this case $R_{oe} = Q\rho$ was ~ 5 MOhm ($\rho$ - is the wave resistance of the waveguide) and was not temperature dependent.

The resonator 1 was mounted horizontally in the cell filled with liquid helium and supported by spring-loaded metallic rods set at the disc center. In the whispering gallery mode the rods had a minimum effect on the resonance properties of the resonator and the resonance oscillation energy was concentrated near the cylindrical surface. The resonator was operated in the mode of the passing electromagnetic wave. Dielectric rectangular waveguides 2 made of isotropic quartz served as exciting and receiving antennas. They had an absorbing element at one end. The other rectangular ends were pasted hermetically into hollow metallic pipes (waveguides 3) coming out from the cryostat. The vector of the electric component of exciting electromagnetic field was perpendicular to the disc axis.

The oscillation spectrum of the system consisting of radial and azimuthal modes was registered on-line on monitor 8. Only the azimuthal $HE_{mn\delta}$ - type (whispering gallery) modes with the radial $n = 1$ and azimuthal $m = 17 – 110$ indexes were used in these experiments. The frequency spacing between the adjacent modes was ~ 2 GHz. The frequency tuning was realized by mating block 7. The tuning step was 1 – 10 MHz on rough tuning and 140 kHz on fine tuning. The scanning velocity in the precision regime was 100 Hz/sec. To stabilize the frequency regime, the generator was first subjected to long-duration heating at the required control voltage. In the course of the experiment the control voltage was varied within several millivolts to keep the thermal condition of the backward-wave tube undisturbed.

The resonator was excited with a set of generators 4 covering the frequency range 40 – 200 GHz. Its lower bound was limited by the waveguide cut-off-frequency. Two regime were used – frequency scanning (regime 1) and fixed-frequency measurement (regime 2). The regimes were put into operation with switch 5. The basic measurements were made in regime 1. The control of generators 4 data acquisition and processing were carried out by using receiving module 6 and machine block 7 [15]. To increase the signal-to-noise ratio, the electromagnetic wave was modulated with modulator 9 and low-frequency generator 10. The signal that passed through the low temperature part of the unit was detected with detector 11 and fed to low-bandpass amplifier 12.

The resonance field in the dielectric disc resonator was excited solely by traveling waves. Resonance was observed when the phase of the traveling wave coincided on each subsequent passage of the wave along the generating line of the resonator. The electromagnetic field of these oscillations was concentrated in a narrow spatial area adjacent to the cylindrical surface of the resonator. The field decreased exponentially along the radial axis on moving away from the surface (Fig. 2).

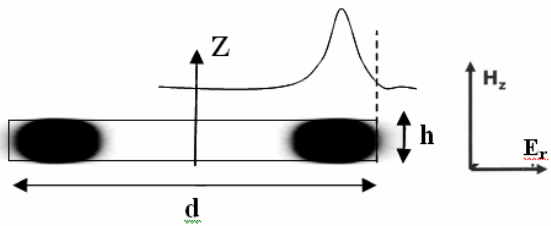

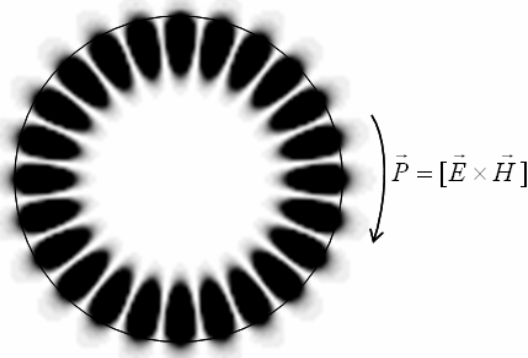

Fig. 2. Electromagnetic energy distribution in the standing wave of whispering gallery mode (m=24) in the dielectric resonator. Dark regions – antinodes of the electric component of the field

## 3. Shapes of resonance peaks.

The amplitude –frequency characteristic of the resonator was first measured in helium gas at T = 4.2 K and the pressure of 3 – 5 Torr. These conditions ensured reliable cooling of the whole system. Then the mode was selected at the required frequency. The highest value of the signal amplitude was taken as a 100 % resonator transmission in this mode. When liquid helium was condensed to the cell, the resonance mode frequencies shifted in accordance with the change in the liquid dielectric permittivity. The signal amplitude decreased as the dielectric losses grew larger. We could thus compare the losses in liquid helium and in vacuum.

Fig. 3 shows the resonance curves for two modes at ~ 67 GHz (Fig. 3*a*) and ~ 180 GHz (Fig. 3*b*). The measurements were made in vacuum (curves on the left) and in liquid helium (curves on the right). The curves were processed by approximating their Lorentz functions (see Ref. [15])

$$A(f) = \frac{A_0}{1 + 4Q^2\left(\frac{f}{f_0} - 1\right)^2}, \quad (2)$$

where A(f) is the current value of the signal amplitude, $A_0$ is the maximum amplitude. Two parameters – resonance frequency $f_0$ and Q-factor – were determined. It is seen that the signal amplitude in liquid helium ($A_L$) is less against the signal amplitude in vacuum ($A_V$).

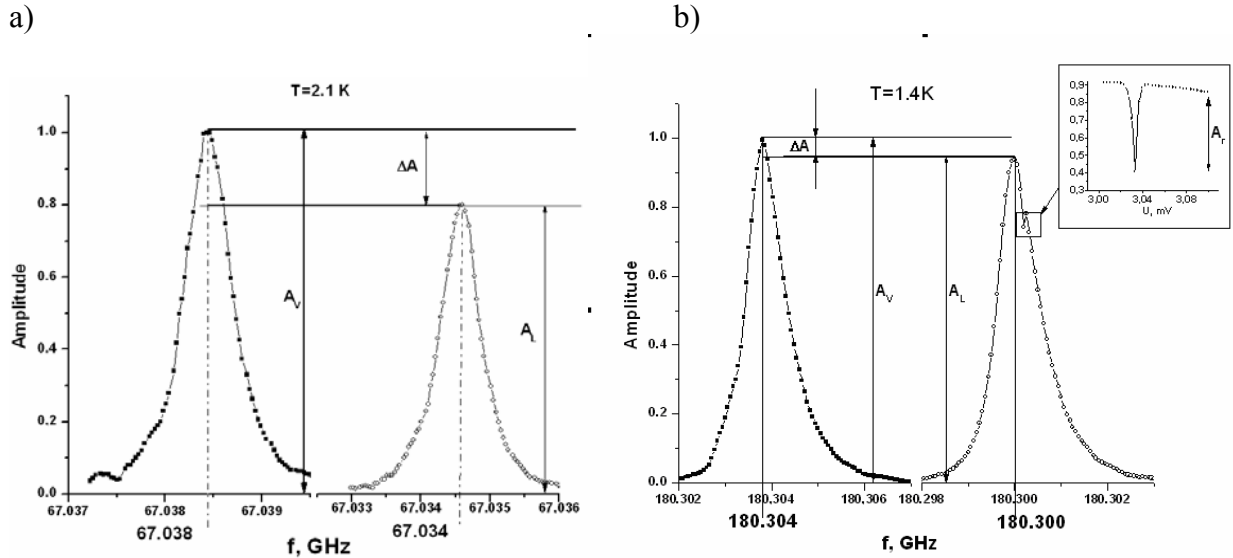

Fig. 3. Typical resonance curves of whispering gallery modes obtained with a dielectric disc resonator with m = 29 (a) and m = 78 (b). • – vacuum at T = 4.2 K;  o - liquid helium. Inset in Fig. b: the portion of resonance absorption in the precision scanning regime (enlarged scale, averaging over 25 curves).

In contrast to other modes, the m = 78 mode had a specific resonance peak in the liquid appearing as a sharp dip in the smooth resonance curve at a certain fixed frequency - dependent on temperature (Fig. 3*b*, inset). We attribute the feature to the resonance absorption of the electromagnetic wave at this frequency (see below). It was found that a decrease in the signal amplitude was frequency-dependent, $\Delta A/A_V = \{A_V - A_L\}/ A_V$. It can be taken as a quantitative characteristic of the relaxation (dielectric) losses in the liquid. The resonance absorption of the electromagnetic wave can be described by the ratio $A_R/A_V$. Since the modes had different amplitudes and Q-factors in vacuum, the $\Delta A$ and $A_R$ values were normalized to $A_V$ of the corresponding mode.

### 4. Creation of single rotons.

The typical spectra of dielectric losses and resonance electromagnetic wave absorption in He II taken at two temperatures are shown in Fig. 4. It is seen that the spectrum of dielectric losses (axis on the left) has a broad maximum in the band of 40 – 80 GHz, which increases with increasing the temperature. The losses can be brought about by consumption of the a.c. electric field energy of the wave to create phonons with energy 2 – 4 K in He II. The process intensity rises on approaching the λ-point (see [10]) presumably due to the fluctuations in the vicinity of the phase transition.

Fig. 4 also illustrates a new effect – resonance absorption of an electromagnetic wave at a certain frequency $f_R$ temperature which depends on temperature. The process has two specific features: (i) the Q-factor of the mode decreases by 5 – 20 % when $f_R$ coincides with the mode frequency; (ii) a very narrow absorption line appears in the smooth spectrum of relaxation losses.

The signal amplitude decreases by $A_R$ at $f = f_R$ (Fig. 3b, inset). Because of the finite bandwidth of the generator, it was impossible to measure the absorption linewidth precisely. The resonance absorption lines were observable owing to their width at the bases (several tens of kHz). The highest power fed to the resonator was about 1 mW. Let note that the effect was power-independent. Namely, the resonance frequency $f_R$ did not change when the power was decreased by ~ 40 dB.

At T = 1.4 K (Fig. 4b) the resonance absorption frequency corresponds to the energy ~ 8.65 K, which coincides with the minimum roton energy $\Delta$ (roton gap). It is natural to relate the effect observed to the creation of roton excitations in He II, which consumes the electromagnetic wave energy. To make sure that scenario was correct, we performed measurements at different temperatures. As $T_\lambda$ was approached, $f_R$ decreased and moved from one mode to another. The temperature dependences of the resonance frequency and the roton gap are shown in Fig. 5 along with neutron scattering data for $\Delta$ [16, 17]. The results obtained by different methods are in very good agreement and thus support the idea of roton creation during the interaction between He II and the electromagnetic wave of a certain frequency.

We should emphasize that the shape of the resonance curve indicated an evolution as the temperature varied (Fig. 6). The resonance curved was measured using the precision regime of scanning (100Hz/s) by changing the oscillator voltage U. The Fig. 6 shows that the resonance absorption lines are resolutionable reliable and their width increases as the temperature rises.

The interaction of electromagnetic waves with elementary excitations in superfluid helium was previously investigated by the Raman method [18]. The Raman spectrum contained a sharp asymmetrical peak corresponding to the energy difference between the incident and reflected photons 18.5 ± 0.5 K, which is close to 2$\Delta$. Later on it was found that the interaction of the electromagnetic wave with He II leads to the formation of a bound double-roton state with the binding energy 0.37 ± 0.10 K [19]. The rotons forming the bound state have almost equal but oppositely directed momentum.

The sharp peak of resonance absorption of the electromagnetic wave observed in our experiments corresponds to creation of a single roton. However, the momentum conservation law does not hold in this partial case because the momentum of the roton exceeds that of an incident photon by many orders of magnitude. Until now, no roton-like quasiparticle with very small momentum has been detected in He II though its possibility has been a subject of discussions [20-22]. In particular, this possibility can be realized when an elementary excitation is produced by transferring the particle from the Bose-condensate to the above-condensate state because this requires finite energy. The spectrum of such excitations has a gap. Note that in its earlier version, the Landau theory of superfluidity admitted along with phonons, the existence of particles whose spectrum has a gap at zero momentum [21].

It is more reasonable to explain the resonance absorption of the electromagnetic wave at f = 180.3 GHz proceeding from the condition of this experiment in which the photon-superfluid helium interaction occurs near the resonator wall. In the spatially homogeneous case the creation of a single roton is impossible at this high frequency because of violation of conservation laws. In our experiment the boundary between the superfluid helium and the dielectric resonator creates inhomogeneity. We can thus assume that the boundary takes on the excessive momentum of the roton. To put it differently, the interaction of the electromagnetic wave with superfluid helium near the resonator wall leads to creation of a phonon in the solid wall [22], as well.

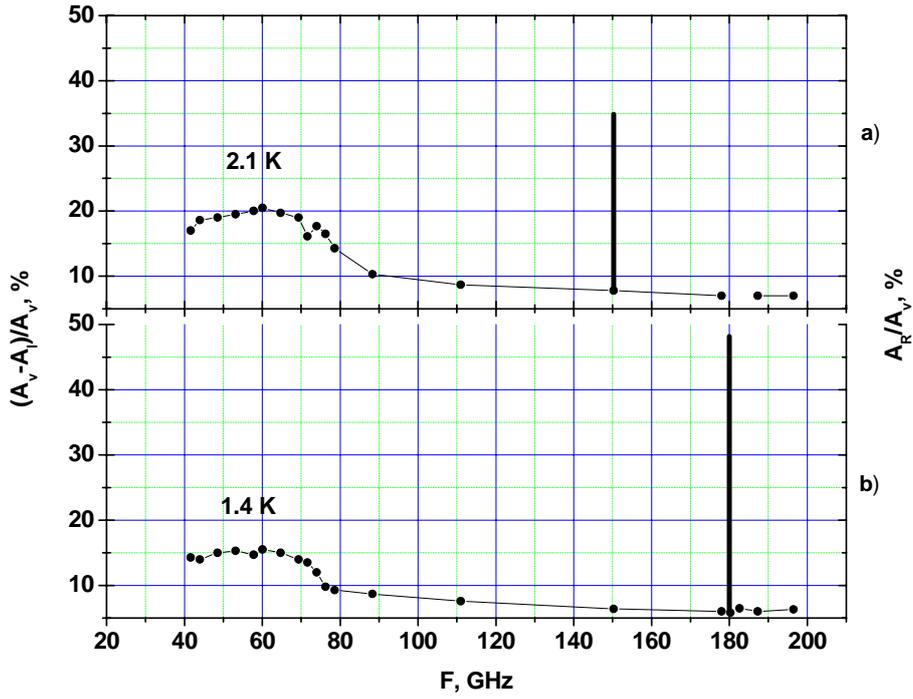

Fig. 4. Frequency dependence of the relative amplitude of electromagnetic wave traveling in He II. a – T = 2.1 K; b – T = 1.4 K. The resonance absorption line is shown by a vertical solid line (the scale of its height is on the right).

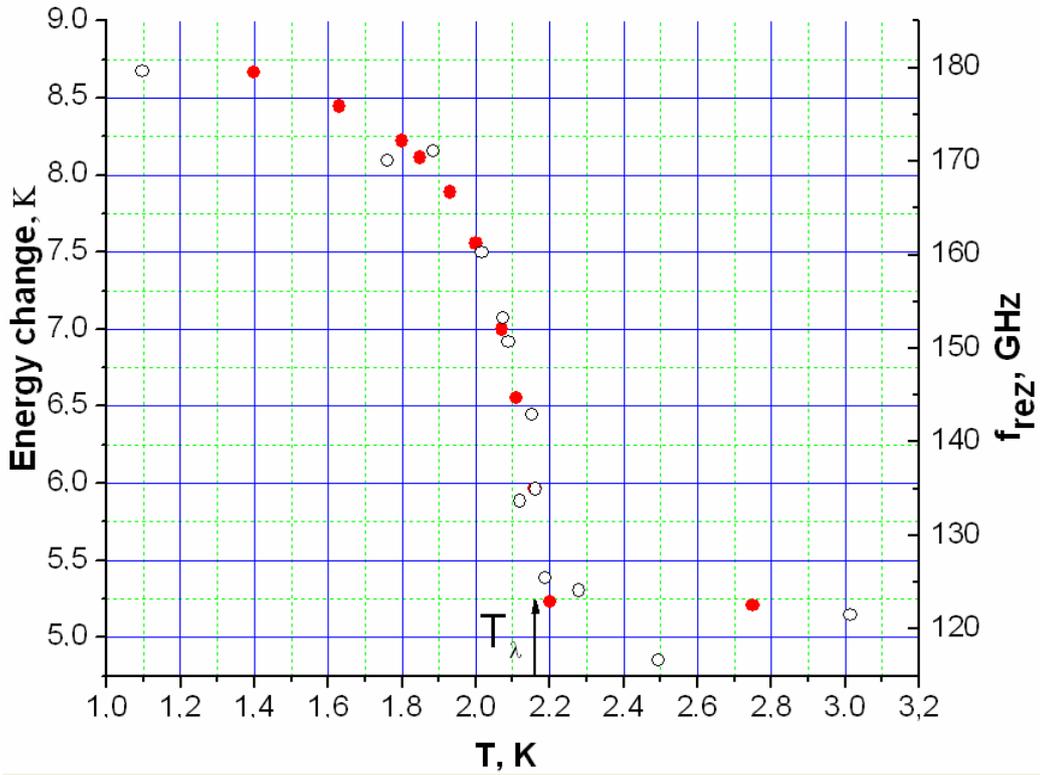

Fig. 5 Temperature dependence of the resonance absorption frequency (right – hand axis) and recalculated roton gap value (left – hand axis):  • - this experiment,  o - neutron experiment [15].

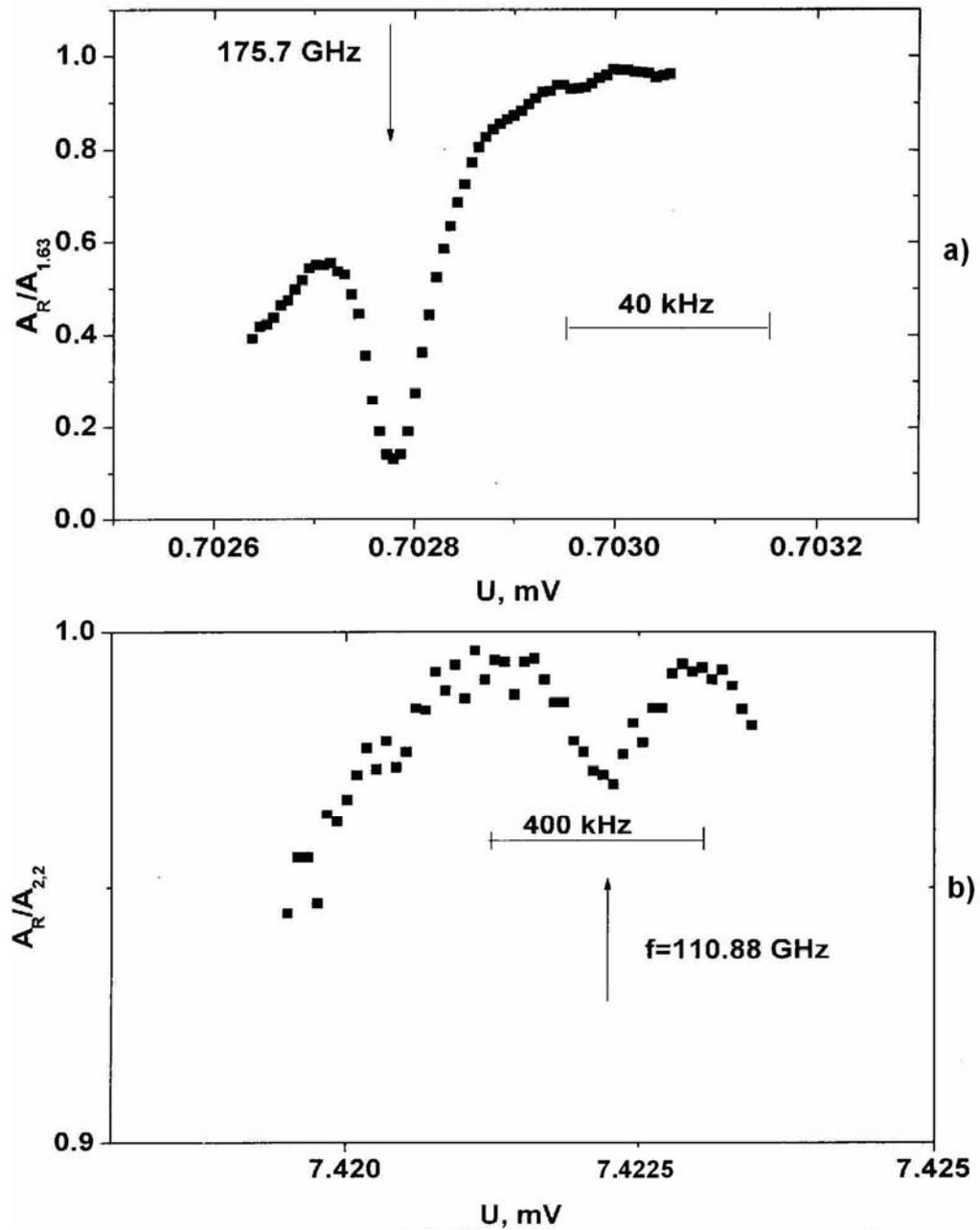

Fig 6. Shape of the resonance curve at various temperatures: a - T=1.63K; b - T=2.2K

The effect of resonance absorption of an electromagnetic wave was also observed above the λ-point. In this case $f_R \sim 120$ GHz, which corresponds to $\Delta \sim 5.2$ K and is in good agreement with neutron scattering data.

A very narrow resonance line means that there are coherent processes in the system. These processes might be associated with interaction between atoms which has an electromagnetic origin. The problem will be studied in the future.

## 5. Kinetics of the signal amplitude variation.

The signal amplitude was investigated under the conditions of varying temperature of liquid helium and its stabilization. The measurements were made in regime 2. The kinetics of the relative amplitude $A_L/A_{2.3}$ on cooling the liquid from 2.3 to 1.9 K and subsequent temperature stabilization ($A_{2.3}$ is the largest signal amplitude at T = 2.3 K) is shown in Fig. 7. The measurement at the frequency of the m = 78 mode shows that absorption increases appreciably as the liquid helium temperature approaches $T_\lambda$. The absorption peak appears several minutes after temperature stabilization (T = 1.9 K). The value of $A_L/A_{2.3}$ takes on a new steady value somewhat higher than that at the initial T = 2.3 K.

The observed features can be interpreted in terms of the following concepts. According to Ref. [1] the ac electric field generates the second sound in He II, i.e. the relative motion of the normal and superfluid components. When liquid helium was cooled in this experiment below the λ-point, the electric component of the electromagnetic wave generated counterflow of the normal and superfluid components around the resonator cylindrical surface. The normal component is clamped near the resonator wall, and persistent superfluid flow is produced. The features of the signal amplitude (see Fig. 7) are due to the interaction of the microwave with the superfluid flow in He II. The character of the electromagnetic wave absorption is strongly dependent on the velocity of the superfluid flow $V_s$. This was demonstrated in special experiments [23] in which $V_s$ was varied using devices similar to Kapitza's "hydrodynamic guns" [24]. The $V_s$ and $V_n$ flows were generated with a heater. Two guns generated superfluid flow moving tangentially to the cylindrical surface of the resonator. As a result, the total superfluid flow could vary from zero to a certain maximum value. The impact of the directed superfluid flows on the electromagnetic wave absorption will be discussed comprehensively in a special paper (see Ref. [23] for brief information and results). Note that the superfluid flow suppresses significantly the Q-factor at all modes, which is indicative of additional dissipation.

The features of the signal amplitude become evident with a certain delay (see Fig. 7) following the superfluid transition. This implies that the formation of superfluid flows takes some time (several minutes). The signal amplitude continued to vary even upon stabilization of temperature. The new regime also took several minutes to reach a steady state. Fig. 7 shows that the state of the system in He II (at 1.9 K) corresponds to larger electromagnetic wave dissipation than in He I (at 2.3 K).

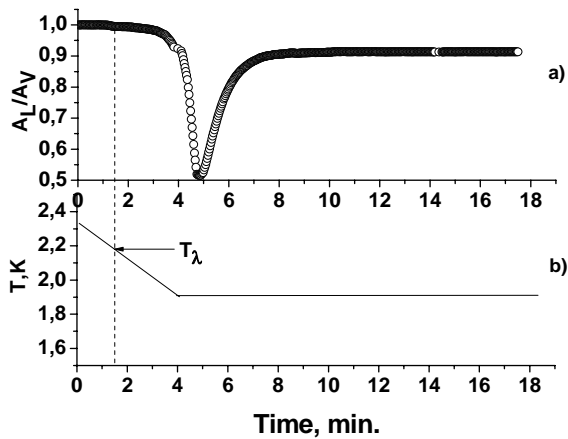
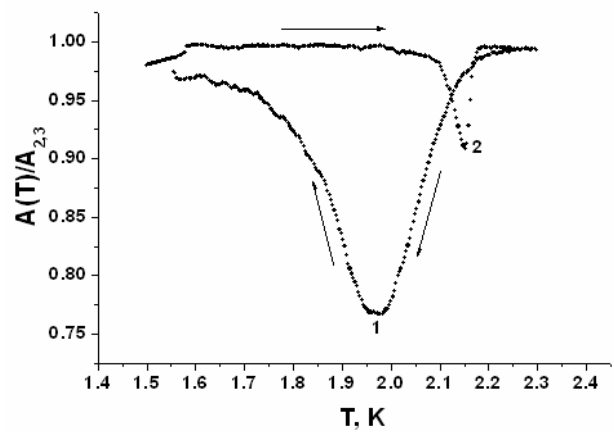

Fig. 7. Time variation of electromagnetic wave amplitude for $A_L(t)/A_V$ (m = 78 mode) at temperatures variation from 2.3 K to 1.9 K and subsequent temperature stabilization - (a); The corresponding time variation of the temperature - (b).

Fig. 8. Temperature dependence of signal amplitude for m = 78 mode on cooling (curve 1) and heating (curve 2) the system.

The results shown in Fig. 6 were obtained when cooling the system. As the system was heated, the temperature dependence of the signal amplitude had a noticeable hysteresis (see Fig. 8). The hysteretic effect may be due to the fact that the superfluid flows formed in He II are quite stable and decay only in the immediate vicinity of the λ-point. A similar hysteretic effect was equally observed in the experiment with a torsion oscillator [2] in which a superfluid flow appeared as well.

Note that earlier the maximum of dielectric losses (f = 36.6 GHz) [10] and a sharp decrease in the Q-factor accompanied by hysteretic effects (f = 73.7 GHz) were also observed for the electromagnetic wave propagated in the liquid helium at temperatures in the vicinity of λ-point.

Another interesting finding is that the superfluid flow around the resonator registered by MW absorption is very sensitive to the cryostat orientation with respect to the Earth's rotation axis. When the cryostat is tilted at a certain angle, the superfluid flow either escapes from the resonator or becomes destroyed. Similar effects have been observed recently by Packard's group [25]. Of course, these results are beyond the scope of this paper and require additional investigations.

### 6. Conclusion.

For the first time we have presented strong evidence that in superfluid helium there is the resonance absorption of microwaves whose frequency corresponds to the frequency of a single roton. This indicates that a MW creates rotons near the resonator wall. The resonance frequency is found to decrease as the temperature rises, which exactly corresponds to the temperature dependence of the roton gap obtained from neutron scattering data. It is important that these results confirm the existence of the relationship between mechanical and electric processes in superfluid helium observed previously. In particular, the electric component of the MW field induces the relative motion of the normal and superfluid components. Since the flow of normal component is clamped near walls, this relative motion reveals itself as a circular superfluid flow around the resonator wall. We assume that the interaction of the MW with superfluid helium occurs owing to liquid polarization and appearance of macroscopic dipole moment whose origin is still to be found. It is quite likely that the superfluid transition is accompanied by a spontaneous ordering of microscopic dipole moments of the atoms.

The authors thank A. F. Andreev, V. N. Grigor'ev, L.Melnikovsky, V. D. Natsik, E. A. Pashitsky, Yu. M. Poluektov, S. M. Ryabchenko, V. N. Samovarov, M. A. Strzhemechny and V. D. Khodusov for helpful discussions.